\documentclass[preprint,aps,pra,showpacs,floatfix,superscriptaddress]{revtex4-1}
\usepackage{graphicx}
\usepackage{feynmp}
\usepackage{amsmath,amsfonts,amssymb,verbatim,float}

\usepackage[usenames]{color}	
\usepackage{colortbl}

\usepackage{ulem} 

\usepackage{hyperref} 
\usepackage{subfigure}
\usepackage{indentfirst}
\definecolor{faded}{gray}{0.45}

\newcommand\rd{\color{red}}  
\newcommand{\bp}{{\bf p}}
\newcommand{\bb}{{\bf b}}
\newcommand{\br}{{\bf r}}

\newcommand{\balpha}{\boldsymbol{\alpha}}

\newcommand{\bk}{{\bf k}}
 
\begin{document}
\thispagestyle{empty}
\title{
 Resonant sequential two-photon ionization of atoms by twisted and plane-wave light
}
\author{V.~P.~Kosheleva}

\affiliation{
Helmholtz Institute Jena, 07743 Jena, Germany
}
\affiliation{
GSI Helmholtzzentrum f\"ur Schwerionenforschung GmbH, 64291 Darmstadt, Germany
}
\affiliation{
Theoretisch-Physicalisches Institut, Friedrich-Schiller-Universit\"at, 07743 Jena, Germany
}
\author{V.~A.~Zaytsev}
\affiliation{
Department of Physics, St. Petersburg State University,
Universitetskaya naberezhnaya 7/9, 199034 St. Petersburg, Russia
}
\author{R.~A.~M\"uller}
\affiliation{
Physikalisch-Technische Bundesanstalt, 
D-38116 Braunschweig, Germany
}
\affiliation{
Technische Universit\"at Braunschweig,
D-38106 Braunschweig, Germany
}
\author{A.~Surzhykov}
\affiliation{
Physikalisch-Technische Bundesanstalt, 
D-38116 Braunschweig, Germany
}
\affiliation{
Technische Universit\"at Braunschweig,
D-38106 Braunschweig, Germany
}
\author{S.~Fritzsche}
\affiliation{
Helmholtz Institute Jena, 07743 Jena, Germany
}
\affiliation{
GSI Helmholtzzentrum f\"ur Schwerionenforschung GmbH, 64291 Darmstadt, Germany
}
\affiliation{
Theoretisch-Physicalisches Institut, Friedrich-Schiller-Universit\"at, 07743 Jena, Germany
}
%
\begin{abstract}
We study the resonant sequential two-photon ionization of neutral atoms by a combination of twisted- and plane-wave light within a fully relativistic framework. 
In particular, the ionization of an isotropic ensemble of neutral sodium atoms ($Z = 11$) from their ground $3^{2}S_{1/2}$ state via the $3^{2}P_{3/2}$ level is considered.
We investigate in details the influence of the kinematic parameters of incoming twisted radiation on the photoelectron angular distribution and the circular dichroism.
Moreover we study the influence of the geometry of the process on these quantities. 
This is performed by changing the propagation directions of the incoming twisted and plane-wave light.
It is found that the dependence on the kinematic parameters of the twisted photon is the strongest if the plane-wave and twisted light beams are perpendicular to each other.
\end{abstract}
%
\maketitle
%
%
%
%
%
%
\section{Introduction}
The interaction of the twisted (or vortex) light beams with the matter has become an important research topic with extensive applications. 
It is explained by the fact that these beams can carry a nonzero projection of the orbital angular momentum (OAM) onto the propagation direction.
This projection, being an additional degree of freedom, provides a unique possibility to get a deeper insight into the role of the OAM in light-matter interactions.
%
%
Moreover, the Poynting vector of the vortex light beams rotates in a corkscrew manner around the propagation direction, and the intensity profile exhibits a ring structure.
Therefore, in processes involving vortex photons, the position and structure of the target play a prominent role, in contrast to the plane-wave case.
%
%
\\ \indent
%
%
Presently, twisted photons are applied in, e. g., nanotechnology~\cite{Soifer}, astronomy~\cite{Lee,Swartzlander}, metrology~\cite{Leyser}, condensed matter physics~\cite{Torres}, and quantum information~\cite{Padgett}.
Most of these and many other applications rely on knowledge about the interaction of twisted light with ions and atoms.
That, in turn, has stimulated investigations of fundamental processes involving twisted light beams and atomic or ionic systems. 
Up to now, theoretical studies of the excitation~\cite{Afanasev_2013_2014_2016,Scholz,Surzhykov_2015,Afanasev_2017_2018,Afanasev_NJP_2018,Schulz_PRA_2019}, ionization~\cite{Mat,Picon,Muller,Seipt,Boning,Paufler,Baghdasaryan,Surzhykov_2016}, and scattering~\cite{Peshkov_PRA97_023802_2018} processes have been presented. 
Experimental investigations were performed, e. g., for the excitation of a single Ca$^{+}$ ion~\cite{Schmiegelow} and the ionization of a gas target consisting of helium atoms~\cite{Kaneyasu}.
For a more in depth discussions of the possible applications utilizing twisted light, see reviews~\cite{Babiker, Knyazev_PU61_449_2018} and references therein.
%
%
\\ \indent
%
%
In the present paper, we perform a fully-relativistic investigation of the resonant sequential two-photon ionization of the alkali-like ions (or atoms) by a combination of plane-wave and twisted light.
We carry out our study on an example of the neutral sodium atoms ($Z = 11$).
For this system, the resonant sequential two-photon ionization proceeds as follows.
In a first step, the plane-wave photon excites the valence electron from $3s$ to $3p_{3/2}$ state, and in the second step, it is ionized by a vortex photon. As a target we consider an isotropic ensemble of atoms as it can be readily performed with the present day techniques. In this case we show that both the photoelectron angular distribution and circular dichroism depend on the ratio of the transversal and longitudinal components of the momentum of the twisted light, which is defined by the so-called opening angle.
We propose to enhance this dependency through an appropriate choice of the geometry, i. e. via adjusting the angle between the plane-wave and twisted photons.
It is found that the sensitivity to the opening angle is the strongest if the incident (plane-wave and twisted) beams are perpendicular to each other.
Moreover, this dependency is stronger than in the case of single-photon ionization by twisted light beams~\cite{Knyazev_PU61_449_2018}. 
%
%
\\ \indent
%
%
The paper is organized as follows:
In Sec.~\ref{sec:pw} the basic equations for the resonant sequential two-photon ionization by plane-wave light are briefly recalled.
In Sec.~\ref{sec:tw} the theoretical description of this process involving a combination of twisted- and plane-wave light is presented.
In Sec.~\ref{sec_rd} we investigate the angular distribution and the circular dichroism for different opening angles of the twisted photon.
The dependence of the ``twistedness''-induced effects on the angle between the first, plane-wave and the second, twisted photon is also presented in Sec.~\ref{sec_rd}.
Finally, a summary and an outlook are given in Sec.~\ref{sec:conclusion}. 
%
%
\\ \indent
%
%
Relativistic units, $m_e = \hbar = c = 1$, and the Heaviside charge unit $e^2 = 4\pi\alpha$ (where $\alpha$ is the fine-structure constant) are used in the paper.
%
%
\section{Basic formalism}
In the present paper, we focus on the investigation of the resonant sequential two-photon ionization of alkali-metal atoms by combination of plane-wave and twisted photons.
The resonant two-photon ionization is a two-step process.
In the first step, the photon excites the target atom from the ground $i$ state to the excited $d$ one, while the second photon ionizes the atom.
Here, we study the scenario where the first photon is conventional (plane-wave) one and the second photon, coming after a slight time delay, is twisted.
We start with the basic expressions for the conventional resonant two-photon ionization since the formulas for the twisted case can be traced back to their plane-wave counterparts.
%
%
\subsection{Resonant sequential ionization by two plane-wave photons}
\label{sec:pw}
%
%
The probability of two-photon ionization in the resonance approximation (see Ref.~\cite{Shabaev_PR356_119_2002} for further details) is given by
\begin{equation}
\frac{dW^{(\rm pl)}_{\bk_2\lambda_2,\bk_1\lambda_1}}{d\Omega_f}
=  
\frac{p_f \varepsilon_{f}}{\Gamma_d^2}
\frac{4(2\pi)^7}{2j_i+1}\sum_{\mu_fm_i}
\left\vert 
\sum_{m_d}
\tau^{(\rm ion, pl)}_{\bp_f\mu_f; \bk_2\lambda_2, dm_d}
\tau^{(\rm exc)}_{dm_d; \bk_1\lambda_1, im_i}
\right\vert^2,
\label{eq_W_pl}
\end{equation}
where $\varepsilon_f$ and $\bp_f$ are the energy and asymptotic momentum of the emitted electron, respectively, $p_f = |\bp_f|$, $j_i$ is the total angular momentum (TAM) of the initial $i$ state, $m_d$ is the TAM projection of the excited $d$ state, and $\Gamma_d$ is the total width of this state.
The first photon is characterized by its momentum $\bk_1$ and helicity $\lambda_1$, and the second photon by $\bk_2$ and $\lambda_2$, accordingly.
In Eq.~\eqref{eq_W_pl}, averaging over the TAM projection of the initial state $m_i$ and the summation over the helicity of the outgoing electron $\mu_f$ are performed.
In the present work, we consider alkali-metal atoms. Therefore, we can apply the single active electron approximation for the description of the resonant two-photon ionization.
In the framework of this approximation, the excitation amplitude is given by
\begin{equation}
\tau^{(\rm exc)}_{dm_d; \bk_1\lambda_1, im_i} = 
- \int d\br \Psi^\dagger_{dm_d}(\br) R_{\bk_1\lambda_1}^{(\rm pl)}(\br) \Psi_{im_i}(\br).
\label{eq_tau_exc}
\end{equation}
Here $ \Psi_{im_i}(\br)$ and $\Psi_{dm_d}(\br)$ are the wave functions of initial and intermediate atomic states, respectively,  $R_{\bk\lambda}^{(\rm pl)} = -e\balpha\cdot{\bf A}^{\rm (pl)}_{\bk\lambda}$ is the photon absorption operator in Coulomb gauge with the vector of Dirac matrices $\balpha$ and the vector potential of the plane-wave photon:
\begin{equation}
{\bf A}^{(\rm pl)}_{\bk\lambda}(\br)
 = 
\frac{\boldsymbol\epsilon_{\bk \lambda}  
e^{i\bk\cdot\br}}{\sqrt{2\omega(2\pi)^{3}} }.
\label{eq_A_pl}
\end{equation}
The ionization amplitude is given by
\begin{equation}
\tau^{(\rm ion,pl)}_{\bp_f\mu_f; \bk_2\lambda_2, dm_d} = 
- \int d\br \Psi^{(-)\dagger}_{\bp_f\mu_f}(\br) R_{\bk_2\lambda_2}^{(\rm pl)}(\br) \Psi_{dm_d}(\br),
\label{eq_tau_ion_pl}
\end{equation}
where $\Psi^{(-)}_{\bp_f\mu_f}$ is the wave function of the outgoing electron whose explicit form can be found, e.g., in Refs.~\cite{Rose,Pratt,Eichler}.
The amplitudes \eqref{eq_tau_exc} and \eqref{eq_tau_ion_pl} uniquely define the probability \eqref{eq_W_pl} and, thus, all the properties of the resonant two-photon ionization process.
%
%
\subsection{ Resonant sequential ionization by plane-wave and twisted photons}
\label{sec:tw}
\begin{figure}[H]
\begin{center}
\includegraphics[width=1\linewidth]{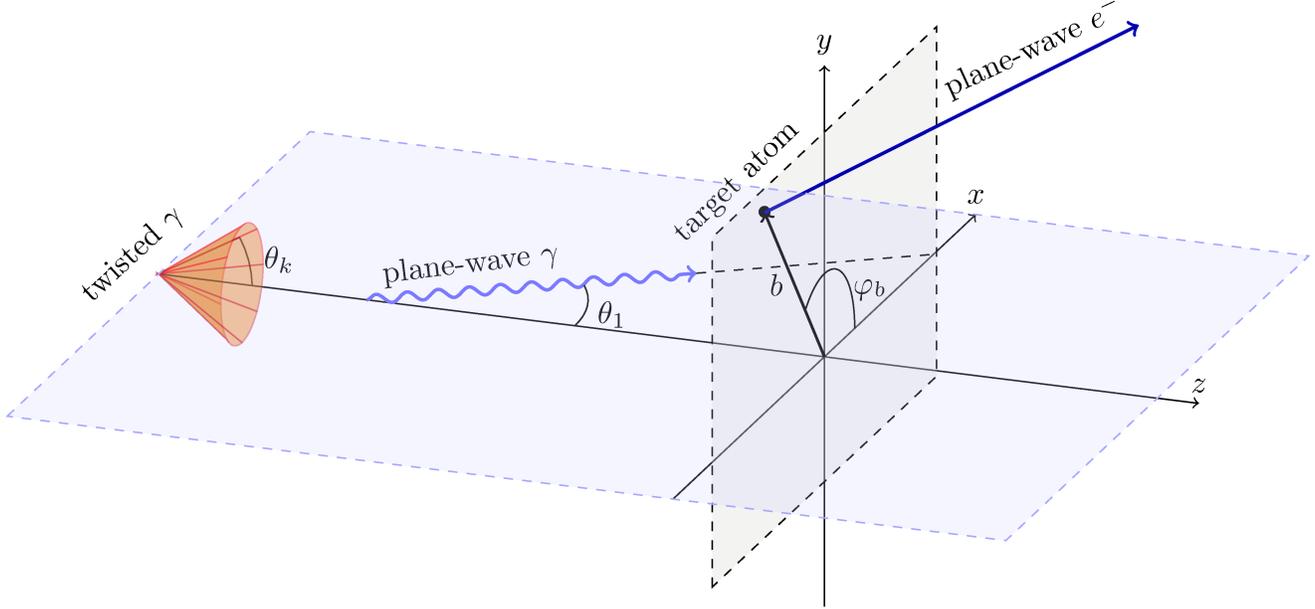}
\caption{Geometry of the resonant sequential ionization of a single atom 
by the combination of plane-wave and twisted light. 
}
\label{ris_geometry}
\end{center}
\end{figure}

%
%
In the present work, we restrict ourselves to the case of Bessel-wave twisted photons.
These waves possess a well-defined energy $\omega$, helicity $\lambda$ as well as projections of the linear $k_z$ and total angular $m$ momenta onto the propagation direction. 
We fix the $z$ axis along this direction. 
The Bessel-wave twisted photon is described by the vector potential~\cite{Mat,Jentschura,Ivanov_PRA84_033804_2011}:
\begin{equation}
{\bf A}^{(\rm tw)}_{\varkappa m k_z \lambda}(\br)
= 
i^{\lambda - m}
\int
\frac{e^{im\varphi_{k}}}{2\pi k_{\perp} }
\delta(k_{\perp} - \varkappa)
\delta(k_{\parallel} - k_{z})
{\bf A}^{(\rm pl)}_{\bk\lambda}(\br)
d\bk,
\label{eq_A_tw}
\end{equation}
where $k_\parallel$ and $k_\perp$ are the longitudinal and transversal components of the momentum $\bk$, respectively, and $\varkappa = \sqrt{\omega^2 - k_z^2}$ is the well defined transversal momentum of the Bessel photon.
From Eq.~\eqref{eq_A_tw}, it is seen that in momentum space, Bessel states represent a cone with the opening angle $\theta_k = \arctan(\varkappa/k_z)$.
In coordinate space, the intensity profile and the flux density of the twisted photon are not homogeneous functions of the space variables (see, e.g., \cite{Mat} for further details).
This distinguishing feature makes the ionization amplitude dependent on the relative position of the twisted photon and the target.
%
%
\\ \indent
%
%
To investigate the resonant two-photon ionization of a single alkali-metal atom by a combination of plane-wave and Bessel photons, we first need to discuss the geometry of this process which is schematically depicted in Fig.~\ref{ris_geometry}.
As was mentioned before, the $z$ axis is directed along the propagation direction of the second, Bessel photon.
The reaction $x$-$z$ plane is formed by $z$ axis and wave vector of the first, plane-wave photon $\bk_1$.
The position of the target atom is given by the impact parameter $\bb = (b, \varphi_b, 0)$ in cylindrical coordinates.
The probability of the process under investigation in the resonance approximation is given by
\begin{equation}
\frac{dW^{(\rm tw)}_{\varkappa m k_z\lambda_2, \bk_1\lambda_1}}{d\Omega_f}(\bb)
= 
\frac{p_f \varepsilon_{f}}{\Gamma_d^2}
\frac{4(2\pi)^7}{2j_i+1}\sum_{\mu_fm_i}
\left\vert 
\sum_{m_d}
\tau^{(\rm ion, tw)}_{\bp_f\mu_f; \varkappa m k_z \lambda_2, dm_d}(\bb)
\
\tau^{(\rm exc)}_{dm_d; \bk_1\lambda_1, im_i}
\right\vert^2,
\label{eq_probability_ion}
\end{equation}
where the amplitude of the ionization by the twisted photon has the following form:
\begin{equation}
\tau^{(\rm ion,tw)}_{\bp_f\mu_f; \varkappa m k_z \lambda_2, dm_d}(\bb) 
 = 
- \int d\br \Psi^{(-)\dagger}_{p_f\mu_f}(\br-\bb) 
R_{\varkappa m k_z \lambda_2}^{(\rm tw)}(\br)
\Psi_{dm_d}(\br-\bb).
\label{eq_tau_ion_tw}
\end{equation}
Here $R_{\varkappa m k_z \lambda}^{(\rm tw)}$ is the photon absorption operator:
\begin{equation}
R_{\varkappa m k_z \lambda}^{(\rm tw)} = -e\balpha\cdot{\bf A}^{\rm (tw)}_{\varkappa m k_z \lambda}.
\label{eq_R_tw}
\end{equation}
Substituting Eq.~\eqref{eq_A_tw} into Eq.~\eqref{eq_tau_ion_tw} and changing the integration variable
$\br$ to $\br - \bb$, one can express the amplitude of the ionization by a twisted photon through the one obtained in the plane-wave case~\eqref{eq_tau_ion_pl}:
\begin{equation}
\tau^{(\rm ion,tw)}_{\bp_f\mu_f; \varkappa m k_z \lambda_2, dm_d}(\bb) 
 = 
\int
\frac{e^{im\varphi_{k}}}{2\pi k_{\perp} }
i^{\lambda - m}
\delta(k_{\perp} - \varkappa)\
\delta(k_{\parallel} - k_{z})
e^{i \bk \cdot \bb} \
\tau^{(\rm ion,pl)}_{\bp_f\mu_f; \bk\lambda_2, dm_d}
d\bk.
\label{eq_tau_ion_tw2}
\end{equation}
%
%
So far we have discussed the resonant sequential ionization of a single atom by the combination of the plane-wave and Bessel photons.
This process is interesting from a theoretical viewpoint but most photoionization experiments deal with extended (macroscopic) targets. 
Therefore, we focus below on the macroscopic target, and we describe such a target as an incoherent superposition of atoms randomly and homogeneously distributed.
The probability of resonant sequential two-photon ionization in this case is given by
\begin{equation}
\frac{dW^{(\rm mac,tw)}_{\varkappa k_z\lambda_2, \bk_1\lambda_1}}{d\Omega_f}
=
\int\dfrac{d\bb}{\pi R^2} \frac{dW^{(\rm tw)}_{\varkappa m k_z\lambda_2, \bk_1\lambda_1}}{d\Omega_f}(\bb)
=
\frac{2}{\pi R \varkappa}
\int_0^{2\pi}\frac{d\varphi_k}{2\pi} 
\frac{dW^{(\rm pl)}_{\bk\lambda_2, \bk_1\lambda_1}}{d\Omega_f},
\label{eq_W_tw_mac}
\end{equation}
where the vector $\bk$ is defined by cylindrical coordinates $(\varkappa, \varphi_k, k_z)$, $1/\pi R^2$ is the cross section area with $R$ being the radius of the cylindrical box.
Note that for macroscopic targets, the probability of the resonant sequential two-photon ionization does not depend on the TAM projection $m$ but is still sensitive to the opening angle of incoming twisted photon.
%
%
\\ \indent
%
%
In the present investigation, we restrict ourselves to the normalized probability
\begin{equation}
\frac{ dW^{(\rm norm, tw)}_{\lambda_2, \lambda_1}}{d\Omega_{f}} 
=
\frac{1}{W^{(\rm avr)}_{\lambda_2, \lambda_1}}
\frac{dW^{(\rm mac,tw)}_{\varkappa k_z\lambda_2, \bk_1\lambda_1}}{d\Omega_f},
\label{eq_macr_norm}
\end{equation}
where
\begin{equation}
W^{(\rm avr)}_{\lambda_2, \lambda_1} = \frac{1}{4\pi}\int d\Omega_{f}
\frac{dW^{(\rm mac,tw)}_{\lambda_2, \lambda_1}}{d\Omega_f}.
\label{eq_macr_aver}
\end{equation}
%
%
%
\section{Results and discussions}
\label{sec_rd}
%
%
Let us proceed to the investigation of the effects of the ``twistedness'' and explore the possibilities of their enhancement.
Since in the present study we consider only the scenario of the macroscopic target these effects constitute in the dependence of the measurable quantities on the opening angle of the vortex light beam.
In the present work we consider two-photon ionization of a valence electron of neutral sodium ($Z = 11$). 
We describe this process within the framework of the single active electron approximation.
The active electron in the ground $3s$, excited $3p_{3/2}$, and continuum states is described by the wave functions being the solutions of the Dirac equation with the effective potential describing the electric field of the nucleus and the spectator electrons.
Here we utilize the so-called X$\alpha$ central potential whose parameters are adjusted in such a way to reproduce the energy of the $3{}^2S_{1/2} - 3{}^2P_{3/2}$ transition, namely $2.10443$ eV~\cite{Sansonetti}.
The energy of the twisted photon $\omega_2 = 3.67796$ eV is chosen to be the same as in the experiment~\cite{Duncanson} where the resonant two-photon ionization of the neutral sodium atoms by two plane-wave photons was studied.
%
%
The radial Dirac equation with the effective potential is solved by the modified RADIAL package~\cite{Salvat1995}.
%
%
\subsection{Ionization probability}
\begin{figure} [H]
\begin{center}
\includegraphics[width=0.81\linewidth]{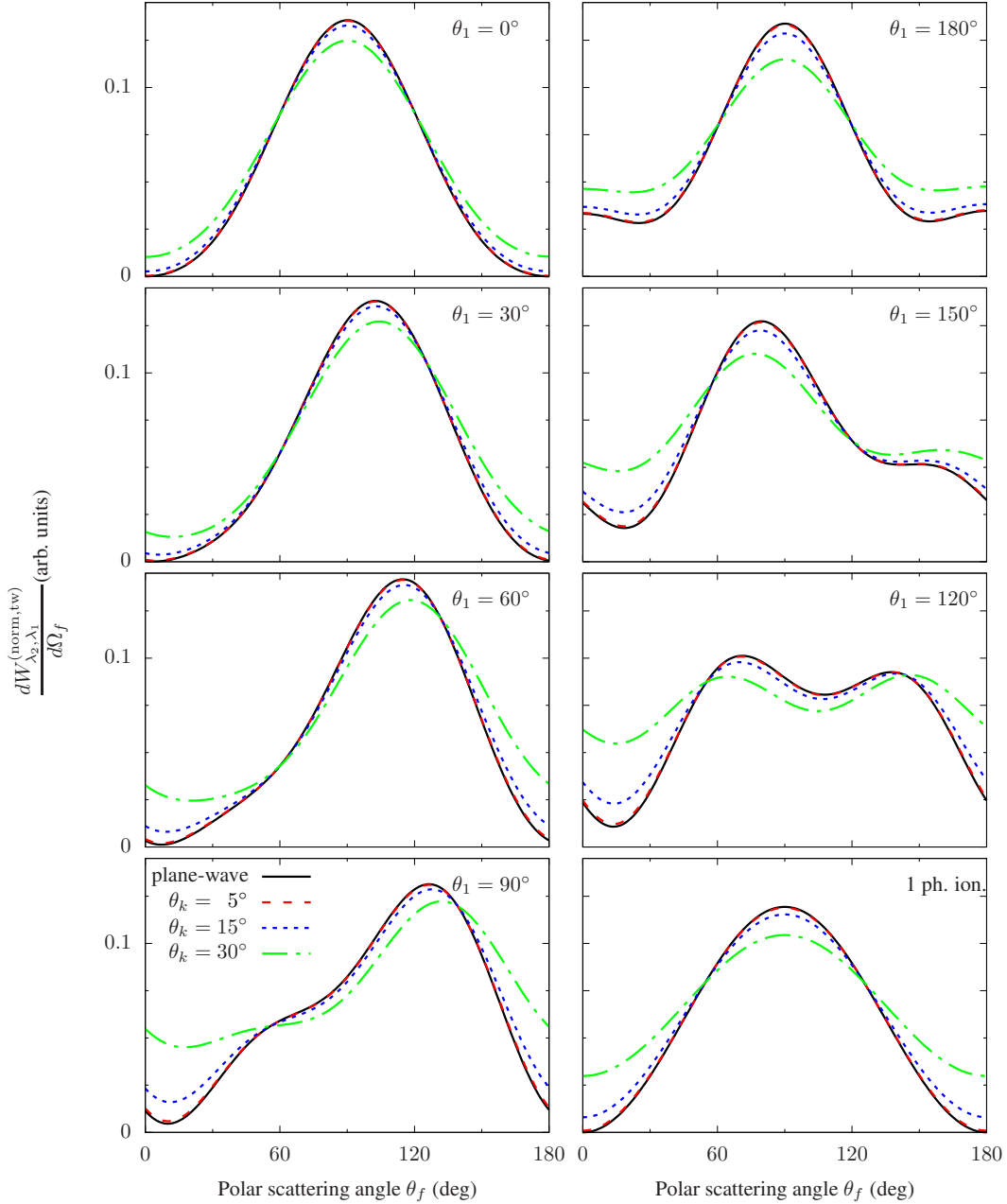}
\caption{
Except the right panel of the last row, the normalized probability~\eqref{eq_macr_norm} of the resonant sequential two-photon ionization of macroscopic sodium target as a function of the polar angle of the ionized electron $\theta_f$ is depicted.
The results are presented for different values of the angle between the first, plane-wave and the second, Bessel photon $\theta_1$ (see Fig.~\ref{ris_geometry}).
The Bessel photon is characterized by its opening angle $\theta_k$.
On the right panel of the last row the normalized probability~\eqref{Knyazev_macr_aver} of single-photon ionization by twisted light is depicted.
}
\label{ris:pw_tw}
\end{center}
\end{figure}

%
%
We start with the analysis of the probability of the resonant ionization of the macroscopic sodium target by two photons with $\lambda_1 = \lambda_2 = 1$.
Except for the right panel of the last {\rd row}, Figure~\ref{ris:pw_tw} presents the normalized probability~\eqref{eq_macr_norm} as a function of the polar angle of the ionized electron $\theta_f$ for different values of the angle between the two photons $\theta_1$.
For reference we show the normalized probability of ordinary two-photon ionization (solid black line).
%
From the upper three rows of this figure, it is seen that the ionization probability changes significantly  when the angle $\theta_1$ is changed to $180^\circ - \theta_1$.
This phenomenon can be understood from the two-photon ionization by co- and counter-propagating plane-wave light (first row in Fig.~\ref{ris:pw_tw}).
In this scenario, the process is invariant under rotations around the $z$ axis, and, as a result, the TAM projection onto this axis is conserved.
Therefore, only the following transitions can take place
\begin{eqnarray}
\label{eq_sch0}
\theta_1 = 0^\circ:&&
\quad
\left\vert 3s_{1/2} \; m_i=\pm \dfrac{1}{2} \right\rangle \rightarrow
\left\vert 3p_{3/2} \; m_d= +\dfrac{1}{2}, + \dfrac{3}{2}\right\rangle \rightarrow
\left\vert \bp_f    \; m_f= +\dfrac{3}{2}, + \dfrac{5}{2} \right\rangle,
\\
\label{eq_sch180}
\theta_1 = 180^\circ:&&
\quad
\left\vert 3s_{1/2} \; m_i=\pm \dfrac{1}{2}\right\rangle \rightarrow
\left\vert 3p_{3/2} \; m_d= -\dfrac{1}{2}, - \dfrac{3}{2}\right\rangle \rightarrow
\left\vert \bp_f    \; m_f= \pm \dfrac{1}{2}\right\rangle,
\end{eqnarray}
%
%
where $\kappa_f$ is the Dirac quantum number of the electron in the continuum and $m_f$ is it's TAM projection onto the $z$ axis. The TAM projection of the intermediate state is $m_d = m_i + \lambda_1$ for co-propagating photons and $m_d = m_i - \lambda_1$ for counter-propagating beams.
From the scheme~\eqref{eq_sch0}, it is seen that $s_{1/2}$ and $p_{1/2}$ waves are absent in the final electron state for the scenario with $\theta_1 = 0^\circ$.
This can, in principle, lead to the significant differences between the cases of co- and counter-propagating plane-wave light beams.
Although for other geometries these simple selection rules are not valid, the contribution of the $s_{1/2}$ and $p_{1/2}$ waves still remains suppressed for the scenarios with $\theta_1 < 90^\circ$, that is supported by numerical results.
%
%
%
%
%
\\ \indent
%
%
From Fig.~\ref{ris:pw_tw}, it is also seen that the ionization probability strongly depends on the opening angle $\theta_k$ of the twisted photon.
As an example, in the case of ionization by two co-propagating plane-wave photons (the upper left panel of Fig.~\ref{ris:pw_tw}), the photoelectrons are not emitted under the angles $0^\circ$ and $180^\circ$.
But for the ionization by a combination of plane-wave and twisted light beams the probability of forward and backward emission is nonzero.
From Fig.~\ref{ris:pw_tw} it is also seen that the closer $\theta_1$ to $90^\circ$, the stronger the dependence of the ionization probability on the opening angle of the ionizing twisted photon $\theta_k$ is.
For $\theta_1 = 90^\circ$ the dependence becomes the most pronounced, as seen from the left panel of the last row in Fig.~\ref{ris:pw_tw}.
In this case, the probability changes significantly, in comparison to the plane-wave one, even for relatively small opening angles.
That makes the scenario with $\theta_1 = 90^\circ$ the most promising for the detection of the kinematic effects in the sequential two-photon ionization.
In addition, the dependence of the probability on the opening angle of the ionizing twisted photon is larger for angles $\theta_1 > 90^\circ$ in comparison to the case where $\theta_1 \rightarrow 180^\circ - \theta_1$.
%
%
\\ \indent
%
%
It is instructive to compare our results with ones presented in Ref.~\cite{Knyazev_PU61_449_2018} where the ionization from the ground state of hydrogen-like ions by twisted photon has been considered.
This comparison should serve only as qualitative indication of the enhancement of the kinematic effects.
Within the nonrelativistic formalism, which was used in Ref.~\cite{Knyazev_PU61_449_2018}, the single-photon ionization probability is given by
\begin{equation}
\frac{ dW^{(\rm norm, tw)}_{\rm 1ph}}{d\Omega_{f}} 
=\dfrac{1}{4\pi}\left[1 - P_{2}({\rm cos}\ \theta_{k}) P_{2}({\rm cos}\ \theta_{f})\right].
\label{Knyazev_macr_aver}
\end{equation}
We note that this ionization probability does not depend neither on the nuclear charge $Z$ nor on the energy of the ionizing photon.
As was shown in Ref.~\cite{Zaytsev}, where the time-reversed process of photoionization, namely radiative recombination, was considered in detail, these dependencies are restored when the relativistic formalism is applied.
To the best of our knowledge, a fully relativistic description of photoionization by twisted photons has been not presented in the literature, yet.
On the right panel of the last row of Fig.~\ref{ris:pw_tw}, the normalized probability of a single-photon ionization by twisted light is depicted.
From Fig.~\ref{ris:pw_tw} it is seen that by choosing properly the geometry in the process of the ionization by the combination of twisted- and plane-wave photons one can significantly enhance the kinematic effects in comparison to the single-photon case.
%
%
%
%
%
\subsection{Circular dichroism}
%

\begin{figure} [H]
\begin{center}
\includegraphics[width=1\linewidth]{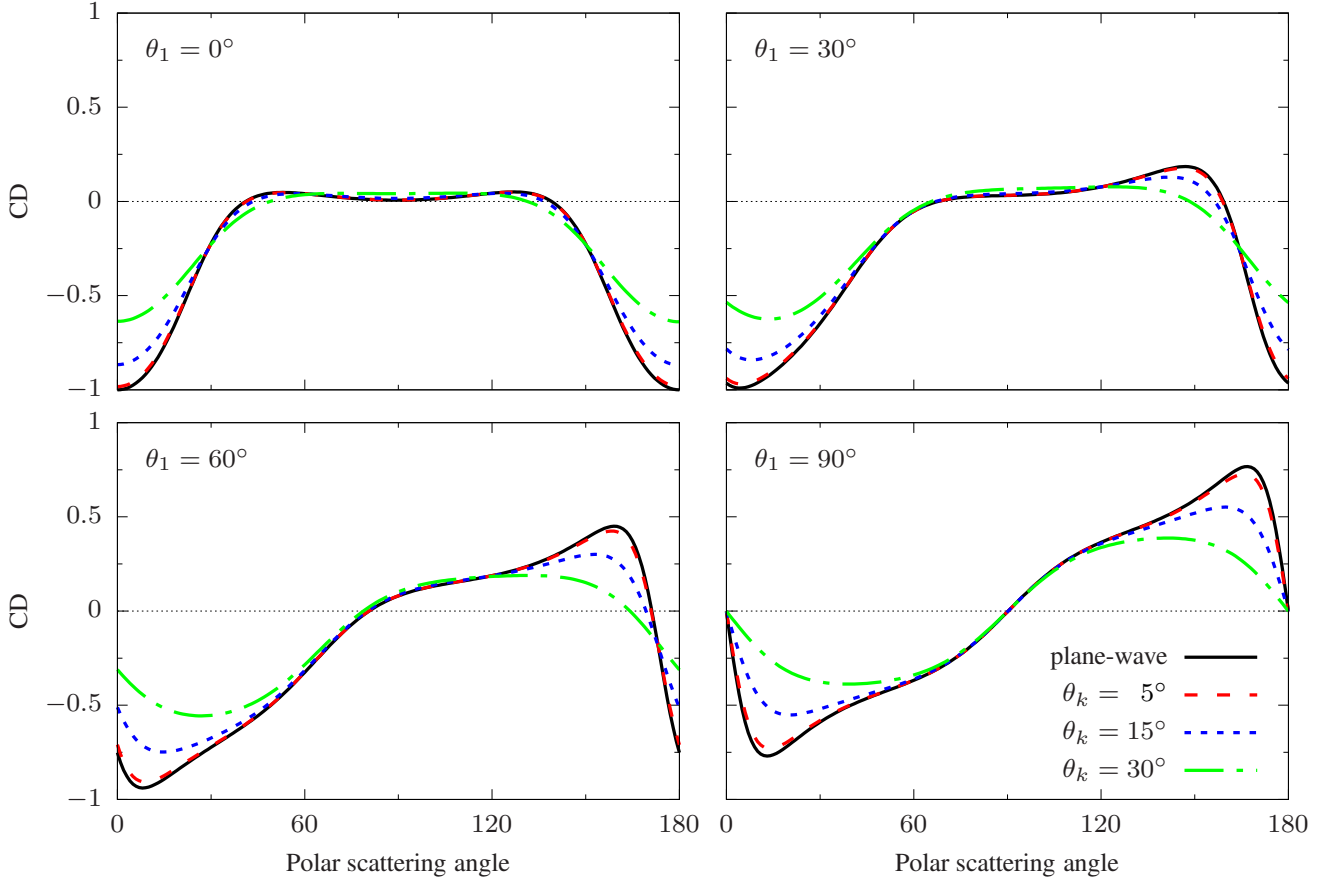}
\caption{
The circular dichroism CD~\eqref{eq_CD} for the resonant sequential two-photon ionization of macroscopic sodium target as a function of the polar angle of the ionized electron $\theta_f$ is shown.
The results are presented for different values of the angle between the first, plane-wave, and the second, Bessel photon, $\theta_1$ (see Fig.~\ref{ris_geometry}).
The Bessel photon is characterized by its opening angle $\theta_k$.
}
\label{ris:dich_pw_tw}
\end{center}
\end{figure}

%
%
The two-photon sequential ionization by a combination of plane-wave and twisted light can be additionally characterized by the dichroism.
As was shown before, for macroscopic targets the angular distribution of photoelectrons does not depend on the TAM projection of the vortex photon. 
Therefore, the dichroism signal in our case can appear only due to a flip of the helicity of the first or second incoming photon. 
Such a signal is commonly known as circular dichroism (CD):
%
%
\begin{equation}
{\rm CD} = \frac{dW_{1,1} - dW_{1,-1}}{dW_{1,1} + dW_{1,-1}},
\label{eq_CD}
\end{equation}
where $dW_{\lambda_2,\lambda_1} \equiv \frac{dW^{(\rm norm, tw)}_{\lambda_2, \lambda_1}}{d\Omega_{f}}$.
Before proceeding to the numerical results for the circular dichroism, we present the following properties of the ionization probabilities
\begin{equation}
dW_{\lambda_2,\lambda_1} = dW_{-\lambda_2,-\lambda_1},
\label{eq_relation}
\end{equation}
and
\begin{equation}
dW_{\lambda_2,\lambda_1}
\xrightarrow[
\theta_f \rightarrow 180^\circ - \theta_f
\atop 
\theta_1 \rightarrow 180^\circ - \theta_1]{}
dW_{-\lambda_2,\lambda_1}.
\label{eq_relation2}
\end{equation}
With the use of these expressions and the results presented in Fig.~\ref{ris:pw_tw} one can calculate the CD.
For the sake of visualization, we present the CD~\eqref{eq_CD} for the scenarios with $\theta_1 < 90^\circ$ in Figure~\ref{ris:dich_pw_tw}.
Let us note, that from the Eqs.~\eqref{eq_relation} and~\eqref{eq_relation2} one can notice that the CD is an antisymmetric function with respect to the simultaneous replacement $\theta_1 \rightarrow 180^\circ - \theta_1$ and $\theta_f \rightarrow 180^\circ - \theta_f$.
Therefore, for $\theta_1 = \theta_f = 90^\circ$ the CD equals zero that can be seen of the right panel of the second row in Fig.~\ref{ris:dich_pw_tw}.
From this graph, one can also see that in the case when the electrons are emitted in the forward or backward directions the CD tends to zero.
One can conclude that in this scenario the ionization probability does not depend on the polarization of the photons.
%
%
In general, from Fig.~\ref{ris:dich_pw_tw} it is seen that, like in the case of the normalized probability, the dependence of the CD on the opening angle $\theta_k$ increase while $\theta_1$ approaches $90^\circ$.
In the case of $\theta_1 = 90^\circ$ the kinematic effects become the most pronounced.
%
%
\section{Conclusion}
\label{sec:conclusion}
%
%
In the present work, we studied the resonant sequential ionization by a combination of plane-wave and twisted light within the fully relativistic formalism.
The study is performed for the ionization of valence $3s$ electron of neutral sodium atom ($Z = 11$) via $3p_{3/2}$ state. We consider a scenario in which the plane-wave and Bessel beam collide with macroscopic target.
In particular, we investigate the photoelectron angular distribution and the circular dichroism for different opening angles of the twisted photon $\theta_k$ and different angles between the first, plane-wave, and the second, Bessel, photons $\theta_1$.
It was found that the kinematic effects do increase with $\theta_1$ approaching $90^\circ$.
That makes the case of $\theta_1 = 90^\circ$ the most promising scenario for the observation of these effects in an experimental realization of the investigated process.
%
%
\\ \indent
We also performed a qualitative comparison between our and the nonrelativistic results presented in Ref.~\cite{Knyazev_PU61_449_2018} for the single-photon ionization from the ground state of hydrogen-like ions by the twisted light.
From this comparison we found that the presence of the plane-wave photon in the two-photon ionization significantly increases the kinematic effects in comparison with ones in the single-photon ionization.
%
%
%
%
%
%
%
%
%
\section*{ACKNOWLEDGMENTS}
%
%
R. A. M. and A. S. acknowledge the support by Deutsche Forschungsgemeinschaft (DFG, German Research Foundation) under Germany's Excellence Strategy -- EXC-2123 QuantumFrontiers -- 390837967. 
%
%
%
%
%

\end{document}